\documentclass[aps,reprint,prb,twocolumn]{revtex4-1}
\pdfoutput=1

\usepackage{graphicx}
\usepackage{amssymb}
\usepackage{amsmath}
\usepackage{amsthm}
\usepackage{slashed}
\usepackage[colorlinks=true, linkcolor=blue, citecolor=blue, urlcolor=blue]{hyperref}
\usepackage{color,rotating}
\usepackage{bbm}
\usepackage{array}
\usepackage{pstricks}
\usepackage{pstricks-add}
\usepackage{colonequals}
\usepackage[utf8]{inputenc}

\DeclareMathAlphabet{\EuRoman}{U}{eur}{m}{n}
\SetMathAlphabet{\EuRoman}{bold}{U}{eur}{b}{n}

 \graphicspath{{./figures/}}

\newcommand{\UV}{{\small UV}}
\newcommand{\IR}{{\small IR}}
\newcommand{\FRG}{{\small FRG}}
\newcommand{\RG}{{\small RG}}

\newcommand{\NLO}{{\small NLO}}

\newcommand{\PMS}{{\small PMS}}

\newcommand{\eg}{{\textit{e.g.}}}
\newcommand{\ie}{{\textit{i.e.}}}

\allowdisplaybreaks

\begin{document}

\title{Ising and Gross-Neveu model in next-to-leading order}

\author{Benjamin Knorr}
\email[Electronic address: ]{benjamin.knorr@uni-jena.de}
\affiliation{Theoretisch-Physikalisches Institut, Universit\"at Jena,
Max-Wien-Platz 1, 07743 Jena, Germany}


\begin{abstract}
We study scalar and chiral fermionic models in next-to-leading order with the help of the
functional renormalisation group. Their critical behaviour is of special interest in
condensed matter systems, in particular graphene. 
To derive the beta functions, we make extensive use of computer algebra.
The resulting flow equations were solved with
pseudo-spectral methods to guarantee high accuracy.
New estimates on critical quantities for both the Ising and the
Gross-Neveu model are provided. For the Ising model, the estimates agree with earlier renormalisation group studies
of the same level of approximation. By contrast, the approximation for the Gross-Neveu model retains many more operators
than all earlier studies. For two Dirac fermions, the results agree with both lattice and large-$N_f$ calculations,
but for a single flavour, different methods disagree quantitatively, and further studies are necessary.
\end{abstract}

\maketitle

\section{Introduction}

Since at least 2010, when the Nobel Prize was awarded for ``groundbreaking experiments regarding
the two-dimensional material graphene'' \cite{GrapheneNP},
the interest in graphene \cite{Novoselov2005, Geim2007, CastroNeto:2009zz, RevModPhys.83.407} and related materials \cite{PhysRevB.76.075131, APL:1.3419932,
PhysRevLett.108.155501, doi:10.1021/nn504451t, PhysRevLett.102.236804, doi:10.1021/nn4009406, doi:10.1021/acs.nanolett.5b00085,
PhysRevLett.101.126804, Guinea5391, PhysRevB.86.081405, Yankowitz2012, Ponomarenko2013, Hunt1427,
SoltanPanahi2011, PhysRevLett.110.106801, PhysRevB.93.134502, PhysRevB.79.241406, Tarruell2012, Polini2013, PhysRevLett.108.140405,
Liu864, Neupane2014, PhysRevLett.113.027603} has grown tremendously. Despite all effort,
the theoretical description of these materials is still difficult. The possibility to theoretically predict
material properties such as conductivity or the band structure would without doubt have a deep impact on many areas.
The relativistic symmetry of the noninteracting low-energy degrees of freedom in graphene gives a strong impetus for a study
of effective field theories on the basis of interacting Dirac fermions \cite{doi:10.1080/00018732.2014.927109, PhysRevB.80.081405,
PhysRevB.86.121402, PhysRevB.66.045108, PhysRevB.75.134512, PhysRevLett.97.146401, PhysRevLett.100.146404, Herbut:2009vu,
PhysRevLett.100.156401, PhysRevB.87.085136, PhysRevB.89.035103, PhysRevB.79.085116, Janssen:2014gea, PhysRevLett.98.186809,
PhysRevB.82.035429, PhysRevB.85.155439, PhysRevB.90.035122, PhysRevB.92.155137, PhysRevLett.106.236805, RevModPhys.84.299,
PhysRevLett.86.958, Gehring:2015vja, PhysRevB.78.165423, PhysRevB.81.125105, 1751-8121-45-38-383001, PhysRevB.81.085105, PhysRevB.87.041401,
PhysRevB.82.205106, PhysRevD.88.065008, Jakovac2015, Scherer:2016zwz}.
A particular model for Dirac materials is given by a combination of two Gross-Neveu type models \cite{Classen:2015mar}.
The present work deals with the Ising-like subset of this model, corresponding to the three-dimensional Gross-Neveu model
for four-component Dirac fermions in a reducible representation. In the context of layered materials, it serves
as a minimal model exhibiting a quantum phase transition to a symmetry broken phase and gap formation as a function of coupling strength.
In a field theory context, the Gross-Neveu model can also be regarded as a toy
model for asymptotic safety \cite{Braun:2010tt}.

Many insights in condensed matter systems and quantum field theories are obtained via perturbative expansions.
Most of these series expansions only converge asymptotically, if at all.
To deal with nonperturbative effects, other methods need to be employed.
It is quite possible that a nonperturbative treatment can even solve some very fundamental problems, for example the inclusion of
a putative quantum gravity into the Standard Model, or explain Dark Matter and Dark Energy.
As not every theory can be simulated on a lattice, we shall focus here on the continuous realisation of the exact
renormalisation group by Wetterich \cite{Wetterich:1992yh}.

From a technical perspective, when one wants to apply functional methods to a given theory, several steps are
necessary. Only rarely is an exact solution possible, thus firstly one has to decide on an approximation (truncation)
that one wants to study. Secondly, one has to determine the renormalisation group flow of all operators
present in the truncation. It is clear that the more operators are kept, the more difficult
to calculate the flow. Finally, the resulting (integro-)differential equations need to be solved. For the latter
part, recently a very efficient possibility to solve such equations globally and to high accuracy
by pseudo-spectral methods was put forward in the present context in Ref. \onlinecite{Borchardt:2015rxa}, and we shall also employ this method here.
In Ref. \onlinecite{Borchardt:2016kco}, the fixed point structure of the $O(N)\oplus O(M)$-model was studied pseudo-spectrally between two and three dimensions.
Lately, the method was extended to the integration of flows \cite{Borchardt:2016pif, Borchardt:2016xju}.

In contrast to the situation in perturbation theory, not much is known about the convergence properties
of approximations to the exact renormalisation group from first principles. One possibility is
of course the inclusion of more and more operators, and checking the convergence of observables, as
critical exponents of a phase transition, see \eg{} Refs. \onlinecite{Canet:2003qd, Heilmann:2014iga} for studies in next-to-next-to-leading order,
and Ref. \onlinecite{Morris:1994ie} for a general discussion of the convergence of the derivative expansion.
Clearly, this gives only circumstantial evidence, and a fundamental
understanding of convergence properties is desirable. Nevertheless, it is one of the few means to
judge the quality of a given approximation, when a comparison with results obtained from other methods is not possible.

The focus of this work is threefold: with the help of a simple Ising and a Gross-Neveu model, we shall point out how
to tackle the problem of deriving the renormalisation group flow equations for a large number of operators with the help of computer algebra.
As a showcase, the flow equations for the Gross-Neveu model
in next-to-leading order in a derivative expansion
are derived, which includes 10 operators. These equations are then solved
with pseudo-spectral methods along the lines of Ref. \onlinecite{Borchardt:2015rxa}, showing that indeed there is in principle no limit
to their applicability. With these results at hand, we finally can estimate the convergence properties of
the derivative expansion in these kinds of models by comparing to results with fewer operators, or
results obtained by different methods.

The paper is structured as follows. In \autoref{sec:FRG}, we collect some basic information on the functional
renormalisation group (\FRG{}), followed by \autoref{sec:model}, where the models that we will discuss are introduced.
Statements on the derivation of the flow equations and the regulator choice are given in \autoref{sec:flows}.
Afterwards, we discuss the results on the Ising model in \autoref{sec:Ising}, and on the Gross-Neveu model in
\autoref{sec:GN}. We conclude with a summary and a short outlook in \autoref{sec:summary}. The appendix
collects our conventions for the Clifford algebra.

\section{Functional renormalisation group}\label{sec:FRG}

The \FRG{} is a widely used nonperturbative tool to investigate
quantum fluctuations in a controlled manner. For this, an effective average action, $\Gamma_k$,
is used, interpolating between the microscopic, or classical action, and the full effective action,
which includes all quantum fluctuations. Its dependence on the renormalisation group scale is
governed by the Wetterich equation \cite{Wetterich:1992yh},
\begin{equation} \label{eq:Wetterich}
 \partial_t \Gamma_k = \frac{\mathbf{i}}{2} \mathrm{STr}\left[ \left(\Gamma^{(2)}_k + R_k \right)^{-1} (\partial_t R_k) \right], \quad t=\log\bigg(\frac{k}{\Lambda}\bigg),
\end{equation}
which is a formally exact functional (integro-)differential equation. In this equation, $\Gamma_k^{(2)}$ denotes the
Hessian of the effective average action, and $t$ is the renormalisation group ``time'', which measures
momenta in orders of magnitude of a fixed momentum scale $\Lambda$. Furthermore, the STr sums over discrete and
integrates over continuous indices, and additionally provides a minus-sign for fermions. Finally, the regulator
$R_k$ acts as a dynamical mass and makes every renormalisation group step well-defined, providing both ultraviolet (\UV{})
and infrared (\IR{}) regularisation. Reviews on the \FRG{} can be found in Refs. \onlinecite{Berges:2000ew,Pawlowski:2005xe,Gies:2006wv,Kopietz:2010zz,Delamotte:2007pf}.

In general, it is hard to solve the full flow equation \eqref{eq:Wetterich} exactly. One possible way to obtain a
systematic approximation to the full solution is the derivative expansion, where only momenta up to a certain power,
but arbitrary field dependences are retained. This type of approximation is mostly used in the context of scalar and
fermionic theories. In this work, we restrict ourselves to an approximation which retains
all operators including up to two derivatives (commonly referred to as next-to-leading order (\NLO{})), and at most two fermions.
A possibility to resolve both field and momentum dependence is the {\small{BMW}} scheme \cite{Benitez:2009xg,Benitez:2011xx}.

\section{The model}\label{sec:model}

The Gross-Neveu model describes the interaction of $N_f$ flavours of massless relativistic fermions with a four-fermion interaction \cite{Gross:1974jv}. In our conventions, the corresponding 
microscopic action in Minkowski space reads \cite{Braun:2010tt}
\begin{equation}
 S^\text{GN} = \int \text{d}^3 x \left( \overline\psi \slashed\partial \psi + \frac{\overline g}{2N_f} \left( \overline\psi \psi \right)^2 \right) \, ,
\end{equation}
where $\psi$ denotes a four-component Dirac spinor in a reducible representation of the Clifford algebra.
With the help of a Hubbard-Stratonovich transformation, this can be reformulated as a partially bosonised theory with action
\begin{equation}
 S^\text{pbGN} = \int \text{d}^3 x \left( \overline \psi \left( \slashed\partial + \overline{h} \, \chi \right) \psi - \frac{N_f}{2} \overline m^2 \chi^2 \right) \, ,
\end{equation}
where $\overline g = \overline h^2/\overline m^2$, and $\chi$ is a real scalar field with the same quantum numbers as $\overline \psi \psi$.
This is our starting point for the Gross-Neveu model. Quantum fluctuations will immediately generate all operators consistent with the symmetries.
As already stated, we will include all operators with at most two fermions and two derivatives.

The purely bosonic part of our ansatz for the action reads
\begin{equation}\label{eq:Bos}
 \Gamma_k^\text{bos} = \int \text{d}^3 x \left( \frac{1}{2} Z_\chi(\rho) (\partial_\mu \chi)^2 - V(\rho) \right) \, ,
\end{equation}
where the potential $V$ and the bosonic wave function renormalisation $Z_\chi$ depend on the field, and we introduced $\rho = \chi^2/2$.
This is the approximation to the Ising model that we deal with in \autoref{sec:Ising}.
For the Gross-Neveu model, we first add to this the kinetic term of the fermions and the standard Yukawa coupling,
\begin{widetext}
\begin{equation}\label{eq:FermKin}
 \Gamma_k^\text{ferm} = \int \text{d}^3 x \left( \frac{1}{2} Z_\psi(\rho) \left(  \overline{\psi} (\mathbbm{1}_{N_f} \otimes \slashed{\partial}) \psi 
 - (\partial^\mu \overline{\psi}) (\mathbbm{1}_{N_f} \otimes \gamma_\mu) \psi \right) + g_\chi(\rho) \chi \overline{\psi} \psi \right) \, ,
\end{equation}
\end{widetext}
where both the wave function renormalisation and the Yukawa coupling again depend on the scalar field.
Finally, all further field-dependent interaction terms with at most two fermions and two derivatives are included,
\begin{widetext}
\begin{eqnarray}\label{eq:GNint}
 \Gamma_k^\text{int} = \int \text{d}^3 x \left[
 \mathbf{i}\, J_\psi(\rho) (\partial^\mu \rho) \overline{\psi} (\mathbbm{1}_{N_f} \otimes \gamma_\mu) \psi + X_1(\rho) \chi (\partial_\mu \overline\psi)(\partial^\mu \psi)
 + \frac{\mathbf i}{2} X_2(\rho) (\partial^\mu \chi) \left(  \overline{\psi} \partial_\mu \psi - (\partial_\mu \overline{\psi}) \psi \right) \right. \notag \\
 + X_3(\rho) (\partial^2 \chi) \overline\psi \psi + \frac{1}{2} X_4(\rho) (\partial^\mu \chi) \left( \overline\psi (\mathbbm{1}_{N_f} \otimes \Sigma_{\mu\nu}) \partial^\nu \psi 
 - (\partial^\nu \overline\psi) (\mathbbm{1}_{N_f} \otimes \Sigma_{\mu\nu}) \psi \right) \notag \\
 \left. + \frac{1}{2} (X_5(\rho)+2X_3'(\rho)) (\partial_\mu \chi)^2 \chi \overline\psi \psi \right] \, .
\end{eqnarray}
\end{widetext}
The \NLO{} ansatz for the Gross-Neveu model is thus
\begin{equation}\label{eq:GNansatz}
 \Gamma_k^\text{GN} = \Gamma_k^\text{bos} + \Gamma_k^\text{ferm} + \Gamma_k^\text{int} \, .
\end{equation}
For convenience, we introduced $[\gamma_\mu,\gamma_\nu]=2 \Sigma_{\mu\nu}$.
Our conventions on the Clifford algebra are collected in the appendix.
The specific linear combination in front of the last term in \eqref{eq:GNint} is for convenience, and simplifies the calculation.
All functions are considered to depend on the renormalisation group scale $k$, and the prefactors are
chosen in such a way that all functions are real for a real Minkowskian action. This is important to allow for a clean
projection onto the respective flow equations. Note that the effective action as written above is considered in Minkowski space.
Only after all algebraic manipulations have been executed, we perform the Wick rotation to Euclidean space to be able to
perform the integration over the loop momentum.
The Wick rotation exactly cancels the factor of $\mathbf{i}$ in \eqref{eq:Wetterich}.
We consider $N_f$ fermion flavours in a 4-dimensional reducible representation of the Clifford algebra.
The symmetries of the Gross-Neveu model are discussed in Ref. \onlinecite{Janssen:2014gea} for the case $N_f=2$,
and in Ref. \onlinecite{Scherer:2012nn} for any $N_f$. For our purpose, only the discrete $\mathbbm Z_2$ reflection symmetry,
\begin{equation}
 \psi \to (\mathbbm{1}_{N_f} \otimes \gamma_2) \psi, \, \overline\psi \to -\overline\psi (\mathbbm{1}_{N_f} \otimes \gamma_2), \, \chi \to -\chi \, ,
\end{equation}
with the spatial momentum reflected across the first axis, is relevant. Our ansatz for the action, \eqref{eq:GNansatz}, comprises all operators
that are compatible with this symmetry and realness of the action at the considered order.

Some remarks on the completeness of the ansatz \eqref{eq:GNansatz} are in order. First, we will neglect operators
which contain $\overline \Psi\, \overline \gamma\, \Psi$, where $\overline \gamma = \gamma_0\gamma_1\gamma_2$. Such operators
correspond to a different order parameter, and break time reversal symmetry \cite{Herbut:2009vu, Classen:2015mar}, we will thus neglect them.
Further, one could form contractions of derivatives, $\gamma_\mu$ and $\Sigma_{\mu\nu}$ with the fully antisymmetric symbol $\epsilon_{\mu\nu\rho}$.
The negligence of both can be justified a posteriori: the explicit calculation shows that no such operator is generated by the ansatz above, at least to \NLO{}.

Earlier work only resolved the field dependence of the potential $V$,
while retaining field-independent wave function renormalisations,
and either field dependent \cite{Vacca:2015nta} or field independent \cite{PhysRevLett.86.958, Hofling:2002hj, Braun:2010tt, Braun:2011pp, Janssen:2014gea, Borchardt:2015rxa, Classen:2015mar} Yukawa coupling $g_\chi$,
thus the present work goes beyond earlier limitations.
A supersymmetric version of this model has been investigated in Ref. \onlinecite{Heilmann:2014iga} in next-to-next-to-leading order. Due to the higher
symmetry, there only 4 independent functions had to be considered. The dimensional dependence of this supersymmetric model at criticality
was recently studied \cite{Hellwig:2015woa} as well.

Critical phenomena of physical systems are described by fixed points, which are characterised by the vanishing of all flows of the dimensionless quantities. For this, renormalised
quantities are introduced. The renormalised fields read
\begin{align}
 \hat\chi &= Z_\chi(\overline\rho)^{1/2} k^{-1/2}\chi \, , \notag \\
 \hat\psi &= Z_\psi(\overline\rho)^{1/2} k^{-1} \psi \, , \notag \\
 \hat{\overline{\psi}} &= Z_\psi(\overline\rho)^{1/2} k^{-1} \overline{\psi} \, .
\end{align}
The renormalised potential and wave function renormalisations are defined as
\begin{align}
 \hat V(\hat\rho) &= k^{-3} V(\rho) \, , \notag \\
 \hat Z_\chi(\hat\rho) &= Z_\chi(\overline\rho)^{-1} Z_\chi(\rho) \, , \notag \\
 \hat Z_\psi(\hat\rho) &= Z_\psi(\overline\rho)^{-1} Z_\psi(\rho) \, ,
\end{align}
and the renormalised interaction terms are
\begin{align}
 \hat g_\chi(\hat\rho) &= Z_\chi(\overline\rho)^{-1/2} Z_\psi(\overline\rho)^{-1} k^{-1/2} g_\chi(\rho) \, , \notag \\
 \hat J_\psi(\hat\rho) &= Z_\chi(\overline\rho)^{-1} Z_\psi(\overline\rho)^{-1} k\, J_\psi(\rho) \, , \notag \\
 \hat X_{1-4}(\hat\rho) &= Z_\chi(\overline\rho)^{-1} Z_\psi(\overline\rho)^{-1} k^{3/2} X_{1-4}(\rho) \, , \notag \\
 \hat X_5(\hat\rho) &= Z_\chi(\overline\rho)^{-3/2} Z_\psi(\overline\rho)^{-1} k^{5/2} X_5(\rho) \, .
\end{align}
Here, $\overline\rho$ is an a priori arbitrary, but fixed field value where we normalise the fields to have
a standard canonical kinetic term. One can \eg{} choose the vacuum expectation value (vev) for this,
or zero.
In principle, physical quantities as critical exponents should not depend on such a choice, although
in approximations, there is a residual dependence.
We will use this freedom to check the stability of our results and to get some measure on
the quality of the truncation.
Let us also define the anomalous dimensions,
\begin{align}
 \eta_\chi &= -\partial_t \ln Z_\chi(\overline\rho) \, , \notag \\
 \eta_\psi &= -\partial_t \ln Z_\psi(\overline\rho) \, ,
\end{align}
which carry the scaling of the wave function renormalisations at the normalisation point.

\section{Flow equations}\label{sec:flows}

In this section, we sketch the derivation of the flow equations for the model \eqref{eq:GNansatz}.
The tremendous amount of algebraic manipulations necessitates the use of computer algebra software. We calculated the flow equations using the
Mathematica package xAct \cite{xActwebpage, 2007CoPhC.177..640M,
  2008CoPhC.179..586M,2008CoPhC.179..597M,
  2014CoPhC.185.1719N}, which was originally designed for gravity calculations.
The basic idea to employ it for the derivation of flow equations for scalar and fermionic systems is to introduce a flat manifold representing spacetime,
and additional structure for the actual field content on top of that.
For $O(N)$-symmetric bosonic fields, one introduces another flat manifold, where indices represent the labels of the $O(N)$ symmetry. On the other hand,
for the fermions one only needs a vector bundle over the spacetime manifold. All fields are then represented as tensor fields over the respective
manifold, with a suitable index structure. In particular, one can deal with the abstract Clifford algebra without specifying the representation,
again by introducing suitably indexed tensors and implementing all possible products. For example, the object $\mathbbm{1}_{N_f} \otimes \gamma_\mu$ would carry three indices: the
spacetime index $\mu$, and two Dirac vector bundle indices, indicating the matrix structure. Since for the model considered here, the first factor is always a trivial unit matrix,
we treat it abstractly and refrain from introducing further flavour indices.
With this structure, any needed diagram can be calculated
straightforwardly in \eg{} momentum space. A minimal working example for the case of an $O(N)$ model in \NLO{} can be found in the notebook which is supplemented.

The actual flow equations for \eqref{eq:GNansatz} are way too large to display them \footnote{A notebook containing them is available from the author upon request.}.
Several checks were done to verify these equations. In the corresponding approximation, they agree with the flow equations already derived earlier \cite{Vacca:2015nta}.
Further, different projection schemes must deliver the same flow equations: as an example, the flow of $g_\chi$ can be taken from \eqref{eq:Wetterich}, projected
onto constant fields and onto $\overline\psi \psi$, or from a fermionic two-point-correlator. Mixed correlators give some combination of derivatives of $g_\chi$,
which were explicitly verified to be consistent.

The resulting flow equations were solved with pseudo-spectral methods, which were systematically put forward in the present context in Ref. \onlinecite{Borchardt:2015rxa}. However,
we will solve the equations only up to a finite value of the field, for several reasons. Due to the anomalous dimensions, the large field asymptotic dependence
of the operators is polynomial, but with a real-valued exponent. This impairs convergence tremendously, and would make it even more difficult to solve the equations.
Restricting to a finite region, experience shows that one always gets exponential convergence. The restriction to a finite region has no impact on the
accuracy of the solution, in particular on critical exponents, as long as the region is large enough. To make this clear, one can consider the flow equations
as initial value problem at vanishing field. Given any initial conditions, typically a local solution exists. Only when one asks for global existence, a quantisation occurs,
reducing the set of all local solutions to a (in our case) finite number of global solutions. The important observation then is that, by using any solution method
which is non-local in the sense that it includes information from several distinct points, one can only converge to a global solution, if the region is large enough.
A priori, it is not clear how large the region has to be.
It is dictated by the position of a movable singularity of the fixed point equations.
Thus, there might be a relation to the spike value in a spike plot in the sense of Refs.
\onlinecite{Morris:1994ie, Codello:2012sc, Vacca:2015nta, Hellwig:2015woa}. In practice, one easily realizes when the maximal field value is not large enough.
Then, any change in the number of coefficients or the seed of the Newton-Raphson iteration to solve the flow equations results in a different solution.
For a recent discussion of the convergence of Taylor expansions in this context, see Ref. \onlinecite{Litim:2016hlb}.

\begin{figure*}
 \includegraphics[width=\textwidth]{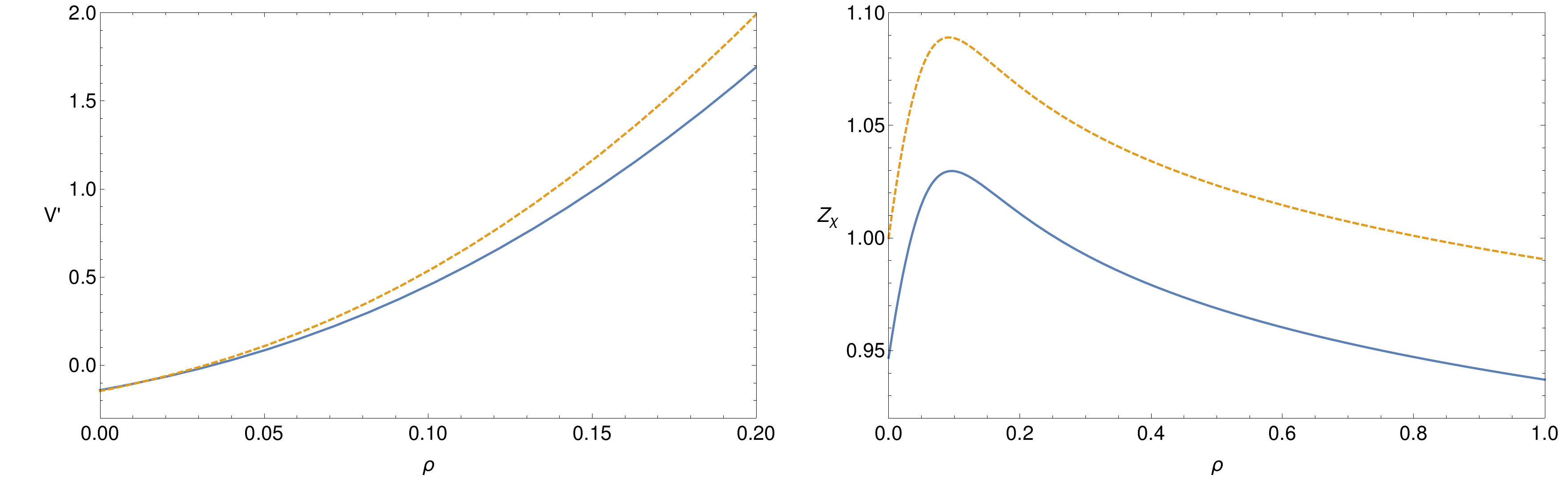}
 \caption{Fixed point solution to the Ising model in \NLO{} with the Litim regulator. The derivative of the effective potential
 is plotted on the left panel, whereas the wave function renormalisation is shown on the right.
 The blue, solid lines correspond to the solution where we fixed $Z_\chi(\rho_0) = 1$, whereas the orange,
 dashed lines have $Z_\chi(0) = 1$.
 One can see that the potentials start to deviate already for small values of the field. On the other hand,
 the two wave function renormalisations differ mainly by a shift.}
 \label{fig:Ising_NLO}
\end{figure*}

Finally, we have to specify the regularisation. For this, the action is amended by
\begin{align}
 \Delta S_\chi &= \int \text{d}^3 x \left( \frac{1}{2} \chi\, R_\chi\left(\frac{p^2}{k^2}\right)\, \chi \right. \notag \\
 &\left.\qquad\qquad+ \overline{\psi}\, R_\psi\left(\frac{p^2}{k^2}\right)\, \frac{\mathbbm{1}_{N_f}\otimes\slashed{\partial}}{p} \psi  \right) \, .
\end{align}
Momentum arguments are to be understood as the momenta after Wick rotation.
As a further quality check of the truncation, we will study several regulator kernels.
A very common choice is the Litim regulator \cite{Litim:2001up},
\begin{align}
 R_\chi(x) &= k^2 (1-x) \, \theta(1-x) \, , \notag \\
 R_\psi(x) &= k \, (1-\sqrt{x}) \, \theta(1-\sqrt{x}) \, ,
\end{align}
where $\theta$ is the Heaviside step function. We will further study the dependence
of the results on the one-parameter family of regulators
\begin{align}
 R_\chi(x) &= \frac{k^2}{2e^{x^a} - 1} \, , \notag \\
 R_\psi(x) &= \frac{k}{2e^{x^a} - 1} \, .
\end{align}
In all cases, for the numerical integration of the threshold functions, an adaptive Gauss-Kronrod 7-15 rule was
employed \cite{9780136272588}. It is enough to consider a finite momentum range due to the regulator insertion in
the flow equation. For the Litim regulator, the integration range is clearly $q\in[0,k]$, on the other hand,
for the class of exponential regulators, the integration range was chosen as
\begin{equation}
 q \in \left[ 0, \sqrt{\frac{6}{5} \left( -\log\left(10^{-b}\right) \right)^{1/a}} \, k \right] \, .
\end{equation}
Here, $b$ is the number of significant digits of the numeric type used, \eg{} 16 for {\textit{double}} precision.
This range is chosen in a way that $\partial_t R$ at the upper limit is always smaller than $10^{-b}$ by several orders of magnitude.

Lastly, a further contribution to some flow equations with the Litim regulator comes
from its distributional character, and has been accounted for analytically. This contribution is unambiguous,
in contrast to the case with a sharp cutoff.

\section{Results for the Ising model}\label{sec:Ising}

Now, we are in the situation to present the solutions to the flow equations. We will start by considering the 
model without fermions.

From now on, we only discuss renormalised quantities, and drop the hats for the sake of readability.
Also, $\rho_0$ denotes the vev, \ie{} $V'(\rho_0)=0$.
In all cases, solutions were computed with pseudo-spectral methods \cite{Borchardt:2015rxa} to at least {\textit{double}} precision.

The Ising model is probably the most studied model within the \FRG{} \cite{Tetradis:1993ts, Morris:1997xj,
Berges:2000ew, Canet:2002gs, Canet:2003qd, Bervillier:2007rc, Benitez:2009xg, Litim:2010tt, Mati:2014xma, Borchardt:2015rxa, Boettcher:2015pja}.
Results with varying levels of truncation exist, which makes a cross-check possible.
We will first discuss the solution obtained with the Litim regulator.

As stated earlier, we compare two possible solutions, where we take the freedom to fix the
renormalised wave function renormalisation to be $1$ at either the vev (denoted by $A$),
or at zero (denoted by $B$). Both the derivative
of the potential and the wave function renormalisation for the two cases are depicted in \autoref{fig:Ising_NLO}.
On the left panel, one can see that the derivative of the potential agrees well up to about $\rho \approx 0.05$.
By contrast, the wave function renormalisations, depicted on the right panel, differ by a constant shift.
The vev and anomalous dimension obtained from the two possibilities are
\begin{align}
 \rho_0^A &= 0.034365 \, , \qquad &\eta^A = 0.050397 \, , \notag \\
 \rho_0^B &= 0.032178 \, , \qquad &\eta^B = 0.049363 \, .
\end{align}

Let us now discuss critical exponents. In general, critical exponents
in the context of the \FRG{} are minus the eigenvalues of the
differential operator obtained from linearising the flow equations around the fixed point.
As we are not interested in the trivial rescaling of the field, additionally we have to demand that
the variation of the wave function renormalisation vanishes at the vev (A) or at zero (B) (in the numerics this shows up
as an exactly marginal eigenvalue).
The first three eigenvalues are
\begin{align}
 \theta_1^A &= 1.59767 \, , \qquad &\theta_1^B &= 1.59660 \, , \notag \\ 
 \theta_2^A &= -0.85878 \, , \qquad &\theta_2^B &= -0.85371 \, , \notag \\
 \theta_3^A &= -1.82275 \, , \qquad &\theta_3^B &= -1.79559 \, .
\end{align}
One sees that due to the different projection scheme, and correspondingly different anomalous dimension,
a similar difference in the critical exponents can be observed. The more irrelevant the exponent, the
higher the difference, indicating that the truncation can only resolve a certain number of them.
For both schemes, the agreement with the literature at the same level of the derivative expansion \cite{Canet:2002gs} is reassuring.

\begin{figure}
 \includegraphics[width=\columnwidth]{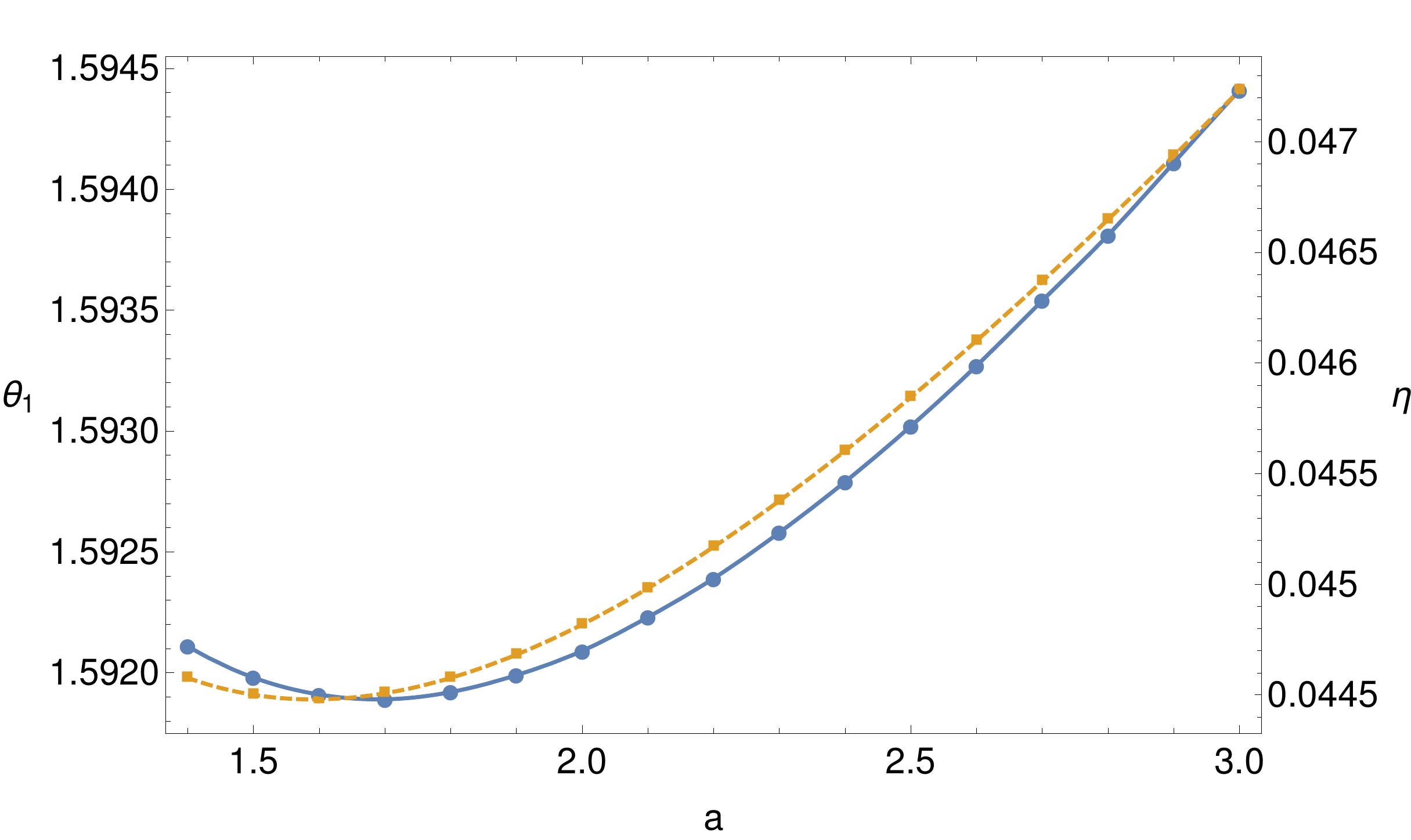}
 \caption{Dependence of the first critical exponent $\theta_1$ (blue dots) and the anomalous dimension
 $\eta$ (orange boxes) of the Ising model on the exponential regulator parameter $a$. An interpolation helps to guide the eye.}
 \label{fig:Ising_NLO_opt}
\end{figure}

Now, we shall present some results obtained with the family of exponential regulators. For definiteness, only the case B
will be discussed. The dependence of $\theta_1$ and $\eta$ on the parameter $a$ is shown in \autoref{fig:Ising_NLO_opt}. The blue dots indicate
values for $\theta_1$, whereas orange boxes stand for values of the anomalous dimension. An interpolation guides the eye.
By the principle of minimum sensitivity (\PMS{}), the extremal values of critical quantities should be closest to the physical values.
In general, this is however not a unique procedure, as different quantities may attain their extremum at different values of the
parameter. We also optimised the second critical exponent, whereas the third critical exponent doesn't show an extremum
in the parameter range considered.
The optimised values of the first two critical exponents and the anomalous dimension, together with their respective optimal $a$, are
\begin{align}
 \theta_1^\text{opt} &= 1.5919 \, , \qquad &a^\text{opt} = 1.69 \, , \notag \\
 \theta_2^\text{opt} &= -0.8479 \, , \qquad &a^\text{opt} = 1.70 \, , \notag \\
 \eta^\text{opt} &= 0.04448 \, , \qquad &a^\text{opt} = 1.59 \, .
\end{align}
These optimised values are much closer to the ``world averages'' over different methods \cite{Pelissetto:2000ek}
\begin{align}
 \theta_1^\text{best} &= 1.587(1)\, , \notag \\
 \theta_2^\text{best} &= -0.84(4)\, , \notag \\
 \eta^\text{best} &= 0.0364(5) \, ,
\end{align}
than the values obtained with the Litim regulator. It is remarkable that the optimal value for $a$ is quite consistent
in optimising the first two critical exponents and the anomalous dimension. Let us finally point out that the second
critical exponent comes out much better than in approximations that don't retain the field dependence of the wave
function renormalisation \cite{Borchardt:2015rxa}, showing that its inclusion improves the quality significantly.

\begin{figure*}
 \includegraphics[width=0.96\textwidth]{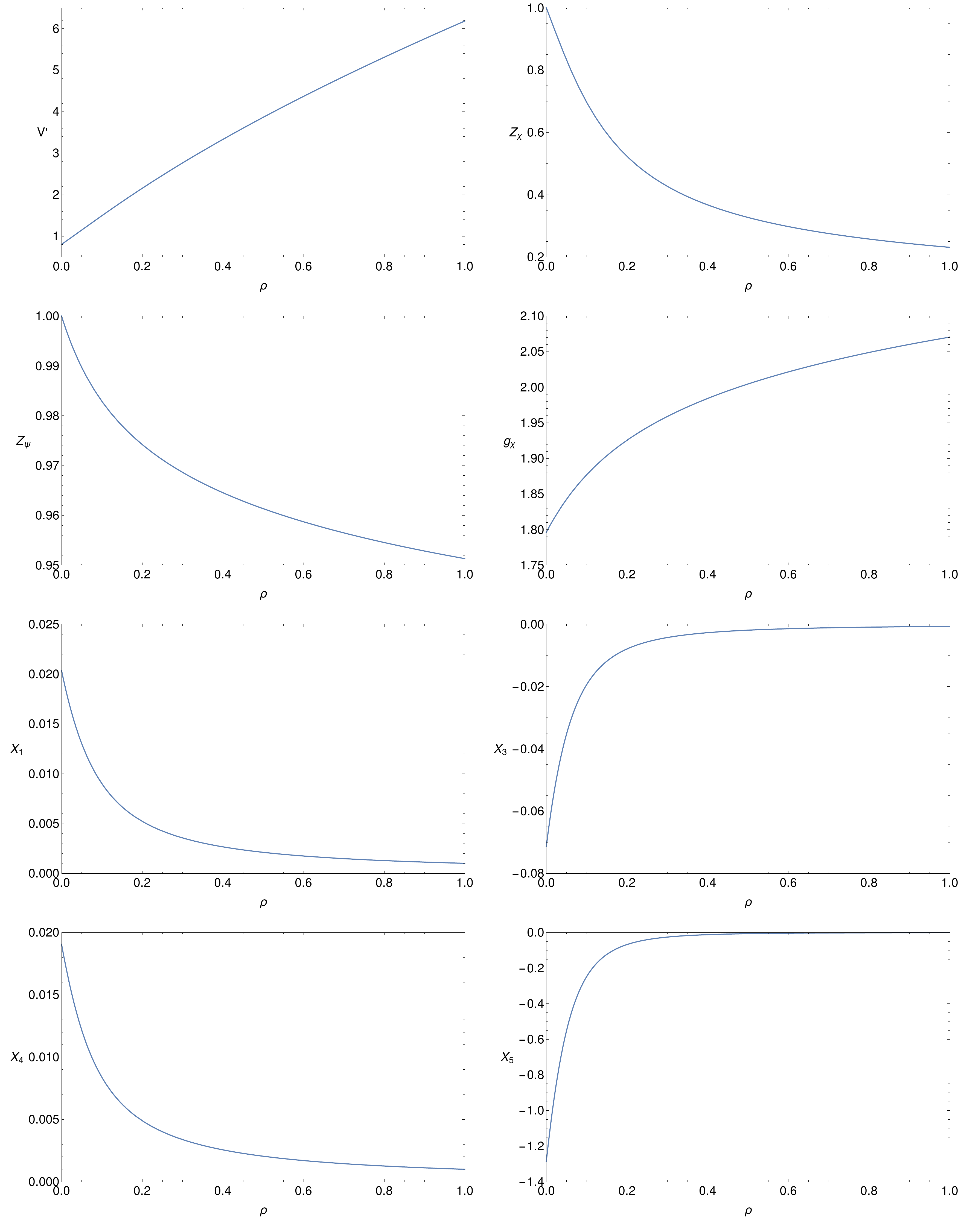}
 \caption{Fixed point solution to the Gross-Neveu model in three dimensions, for two fermion flavours.
 The regulator parameter is $a=2$. Importantly, the function $X_1$ is positive, as it contributes to the denominator of
 propagator functions, and can be seen as the second-order derivative analogue of the usual Yukawa coupling $g_\chi$.}
 \label{fig:GN_NLO_a_2}
\end{figure*}

\begin{figure*}
 \includegraphics[width=\textwidth]{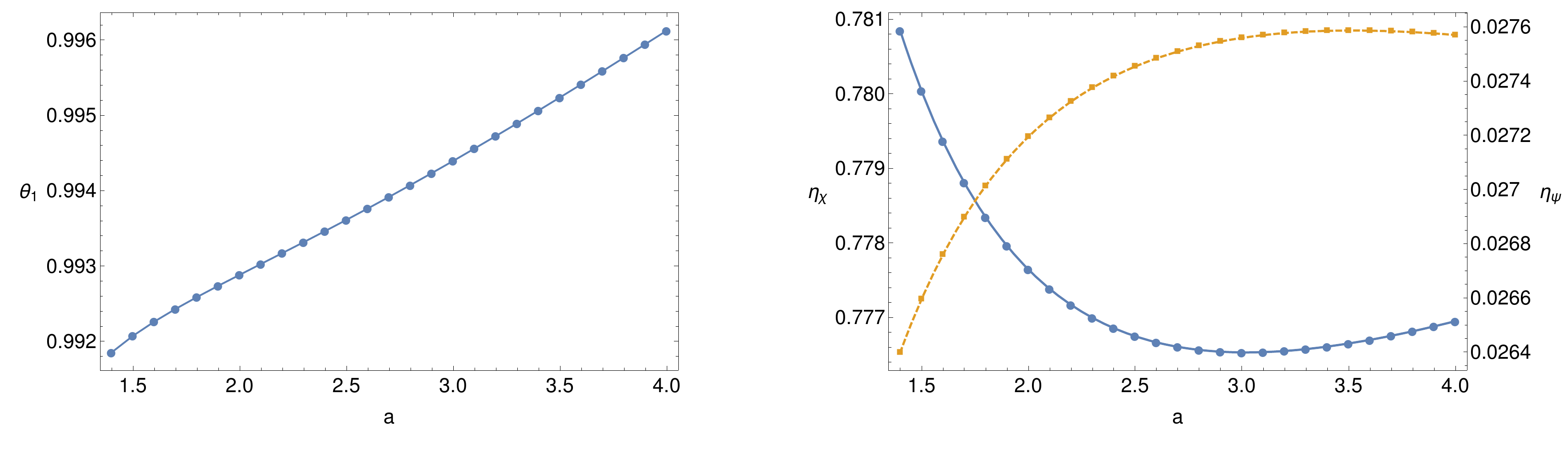}
 \caption{Regulator dependence of physical quantities of the Gross-Neveu model for $N_f=2$.
 On the left panel, the first critical exponent is shown. It doesn't display an extremum, and
 the principle of minimum sensitivity cannot be applied here. On the right panel,
 the bosonic (blue dots) and fermionic (orange boxes) anomalous dimensions are plotted.
 Interpolations help to guide the eye.}
 \label{fig:GN_NLO_Nf_2_regscan}
\end{figure*}

\section{Results for the Gross-Neveu model}\label{sec:GN}

We just saw that the \NLO{} truncation is quite reliable in the Ising model, delivering quantitatively good
results. Now, let us switch to the corresponding model with fermions. We will discuss the cases $N_f=1,2$.
With these parameters, the fixed point is in the symmetric regime, thus there is no difference in the projection schemes A and B.
Further, we shall only discuss the exponential regulator.

In general, due to the quite large bosonic anomalous dimension, the expectation is that improving the
approximation changes the results quantitatively quite a bit. This is only partially the case, as will be seen below.

Before we specialise the fermion flavour number, a general remark is in order. It turns out that the flow equations
of $J_\psi$ and $X_2$ vanish identically if both $J_\psi$ and $X_2$ vanish themselves, \ie{} they have a Gaussian fixed point.
The technical reason for this is subtle, and can be understood easiest in the conventions that we chose. Notice that
the terms with $J_\psi$ and $X_2$ in \eqref{eq:GNansatz} are the only ones which have an explicit factor of $\mathbf i$. Such a factor could only be
generated by the Clifford algebra, but in fact conventions can be chosen such that no explicit factors appear there,
see the appendix. The flow equation \eqref{eq:Wetterich} itself doesn't provide factors of $\mathbf i$ (except the overall prefactor,
which is exactly cancelled by the Wick rotation), as is immanent
when one stays in position space, rather than momentum space. Hence it is indeed expected that these two functions
have a Gaussian fixed point. In systems where one expects a unique fixed point (besides the trivial full Gaussian
and the Wilson-Fisher fixed point), as in our system, such terms can thus be neglected from the outset.

\begin{table}
\begin{tabular}{|c|c|c|c|}
\hline
  & $\theta_1$ & $\eta_\chi$ & $\eta_\psi$ \\ \hline
  FRG (this work) & 0.994(2) & 0.7765 & 0.0276 \\
  FRG \cite{Vacca:2015nta} & 0.996 & 0.789 & 0.031 \\
  Monte-Carlo \cite{Karkkainen:1993ef} & 1.00(4) & 0.754(8) & \textemdash \\
  large-$N_f$ \cite{Gracey:1993kc, Janssen:2014gea} & 0.962 & 0.776 & 0.044 \\
  $(2+\epsilon)$ 3rd order \cite{Gracey:1990sx, Gracey:1991vy, Luperini:1991sv} & 0.764 & 0.602 & 0.081 \\
  $(4-\epsilon)$ 2nd order \cite{Rosenstein:1993zf} & 1.055 & 0.695 & 0.065 \\ \hline
\end{tabular}
\caption{Comparison of the first critical exponent and the anomalous dimensions of the Gross-Neveu model with the literature,
for two fermion flavours, $N_f=2$. Apart from the $\epsilon$-expansions, all methods are in very good agreement.}
\label{tab:litNf2}
\end{table}

We will first discuss the case $N_f=2$. The non-vanishing fixed point functions are shown in \autoref{fig:GN_NLO_a_2}, for the regulator parameter $a=2$.
As expected, the derivative of the potential is strictly positive, indicating that we are in the symmetric regime. The operators
$X_1$, $X_3$ and $X_4$ are parametrically suppressed, as expected from their mass dimension. By contrast, $X_5$ is quite large, but the corresponding
operator comes with $\sim\chi^3$, which suppresses it for small field values.

Again, we study the regulator dependence of the first critical exponent and the anomalous dimensions to optimise the choice of $a$. This dependence is
plotted in \autoref{fig:GN_NLO_Nf_2_regscan}. The optimised values are
\begin{align}
 \eta_\chi^\text{opt} &= 0.7765 \, , \qquad &a^\text{opt} = 3.02 \, , \notag \\
 \eta_\psi^\text{opt} &= 0.0276 \, , \qquad &a^\text{opt} = 3.52 \, .
\end{align}
The first critical exponent doesn't show an extremum, and cannot be optimised by \PMS{}. In the parameter region that was considered, it ranges between $0.992$ and $0.996$, thus we estimate
\begin{equation}
 \theta_1 = 0.994(2) \, .
\end{equation}
In \autoref{tab:litNf2}, we compare to results from the literature. The general agreement of all methods is satisfactory, except for the results coming from
$\epsilon$-expansions. This deviation is not surprising since for the case discussed here, $\epsilon=1$.
In comparison to the former \FRG{} results \cite{Vacca:2015nta}, the critical exponent only changes on the per mille level, depending on the choice of regulator.
The bosonic anomalous dimension changes by $2\%$, the fermionic anomalous dimension by roughly $10\%$.
Let us also mention that recent work \cite{Chandrasekharan:2013aya}
suggested that the compatibility with the cubic-lattice Monte-Carlo results \cite{Karkkainen:1993ef} might be a coincidence, as there a sign problem was ignored, and it is even not clear
if the symmetries in the continuum are the same.

\begin{figure*}
 \includegraphics[width=\textwidth]{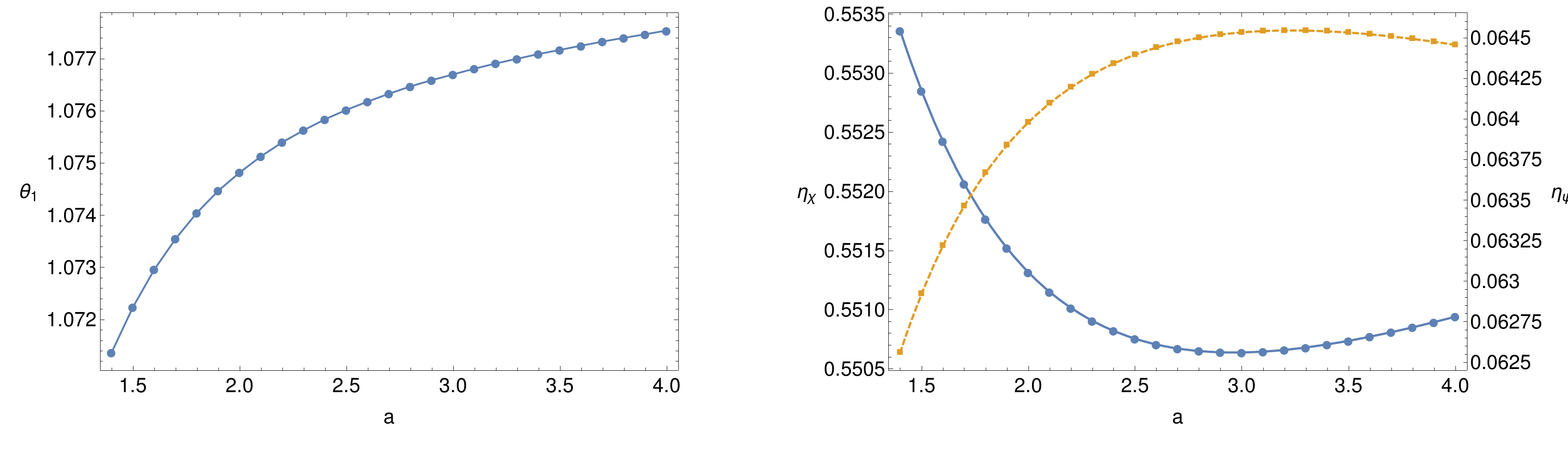}
 \caption{Regulator dependence of physical quantities of the Gross-Neveu model for $N_f=1$.
 On the left panel, the first critical exponent is shown. Similar to the case of two fermion flavours,
 it shows no extremum, and the principle of minimum sensitivity cannot be applied here. On the right panel,
 the bosonic (blue dots) and fermionic (orange boxes) anomalous dimensions are plotted.
 Interpolations help to guide the eye.}
 \label{fig:GN_NLO_Nf_1_regscan}
\end{figure*}

Let us now discuss the case of a single fermion flavour, $N_f=1$. The optimisation with respect to
the regulator is shown in \autoref{fig:GN_NLO_Nf_1_regscan}, and the optimised values are
\begin{align}
 \eta_\chi^\text{opt} &= 0.5506 \, , \qquad &a^\text{opt} = 2.96 \, , \notag \\
 \eta_\psi^\text{opt} &= 0.0645 \, , \qquad &a^\text{opt} = 3.23 \, .
\end{align}
As for the case $N_f=2$, also here the critical exponent doesn't show an extremum, and from the dependence on the parameter $a$ we estimate
\begin{equation}
 \theta_1 = 1.075(4) \, .
\end{equation}
We again compare to different methods in \autoref{tab:litNf1}. In contrast to the case of two fermion flavours, the situation
here is less settled, and different methods don't agree as well. In particular, the results obtained by Monte-Carlo methods \cite{Wang:2014cbw}
deviate significantly from other results. Future work will have to show which results are more trustworthy. When compared to earlier \FRG{} results,
the situation is similar to $N_f=2$: the critical exponent already converged, and only changes by per mille, depending on the regulator. Both anomalous dimensions change by
almost $10\%$.

\begin{table}
\begin{tabular}{|c|c|c|c|}
\hline
  & $\theta_1$ & $\eta_\chi$ & $\eta_\psi$ \\ \hline
  FRG (this work) & 1.075(4) & 0.5506 & 0.0645 \\
  FRG \cite{Vacca:2015nta} & 1.077 & 0.602 & 0.069 \\
  Monte-Carlo \cite{Wang:2014cbw} & 1.25(3) & 0.302(7) & \textemdash \\
  large-$N_f$ \cite{Gracey:1993kb, Gracey:1993kc, Janssen:2014gea} & 1.361 & 0.635 & 0.105 \\
  $(4-\epsilon)$ 2nd order \cite{Rosenstein:1993zf} & 1.160 & 0.502 & 0.110 \\ \hline
\end{tabular}
\caption{Comparison of the first critical exponent and the anomalous dimensions of the Gross-Neveu model with the literature,
for one fermion flavour, $N_f=1$. The Monte-Carlo results conflict with the results obtained by the
other methods.}
\label{tab:litNf1}
\end{table}

\section{Summary and outlook}\label{sec:summary}

This work substantially extends the possible quality of truncations in scalar and fermionic models in functional
\RG{} approaches. On the one hand, one has to calculate the flow equations, which becomes really tedious very quickly if one relies
on by-hand calculation only. By the use of the package xAct \cite{xActwebpage, 2007CoPhC.177..640M, 2008CoPhC.179..586M,2008CoPhC.179..597M,
2014CoPhC.185.1719N}, a truncation with 10 operators could be introduced and the flow
of all operators could be calculated. Second, the resulting complicated non-linear system of differential equations
was solved by pseudo-spectral methods.

Regarding the Ising model, it was shown in \NLO{} that there is some dependence on where one defines the anomalous dimension,
\ie{} where the wave function renormalisation is normalised. This dependence however is quite small, and can be taken as an estimate of systematic
errors. Physical estimates were optimised by a suitable regulator choice. The optimal choice is consistent in optimising both
the first two critical exponents and the anomalous dimension. The second critical exponent comes out much better compared with
approximations only retaining a field-independent wave function renormalisation, which indicates the benefit of resolving this
field dependence.

Concerning the Gross-Neveu model, we also studied the \NLO{} truncation, which includes 10 operators. It turned out that
2 of these operators have a Gaussian fixed point, which can be understood by the fact that they carry explicit factors of
$\mathbf i$ in order to be real, which cannot be generated by the flow. This might point to a hidden symmetry of this model.
For two fermion flavours, the results agree very well with results obtained by very different methods, as Monte-Carlo or large-$N_f$.
For a single fermion flavour, the situation is less satisfactory. \FRG{}, large-$N_f$ and $\epsilon$-expansion
predict a rather large value for the bosonic anomalous dimension, whereas Monte-Carlo methods predict a value which is roughly half
of that. Further research is needed to settle the question at this point.

The present methods can be applied to the case where one considers an $O(3)$ invariant vector coupled to fermions
via Pauli matrices. This is of special interest, as the situation is far less settled across different approaches,
and estimates on critical quantities differ by a large amount \cite{Janssen:2014gea}.
Furthermore, the combination of this model with the model studied
in this work can describe Dirac materials as graphene \cite{Classen:2015mar}. Clearly, both the algebraic as well as the numeric
effort will be considerably higher, however the observation of this work that terms with explicit factors of $\mathbf i$ (with the conventions
as chosen here) can be neglected will help to tackle this problem.

\section*{Acknowledgements}

I would like to thank J. Borchardt, L. Classen, H. Gies, T. Hellwig, S. Lippoldt, S. Rechenberger,
M. M. Scherer, R. Sondenheimer, A. Wipf, L. Zambelli and O. Zanusso for useful discussion during different
stages of this project and H. Gies and A. Wipf for valuable comments on the manuscript.
This work was supported by the DFG-Research Training Group
``Quantum and Gravitational Fields'' GRK 1523/2,
and by the DFG grant no. Wi 777/11.

\appendix

\section{Clifford algebra conventions}

We stick closely to the spin-base invariant formulation \cite{Gies:2013noa,Gies:2015cka,Lippoldt:2015cea}.
Dirac conjugation is defined as
\begin{equation}
 \overline \psi = \psi^\dagger h \, ,
\end{equation}
with an anti-hermitian spin metric $h$. With this choice, the product $\overline \psi \psi$ is real.
Furthermore, we want the kinetic term of the fermions to be real. From this follows that
\begin{equation}
 \gamma_\mu^\dagger = h \gamma_\mu (h^\dagger)^{-1} \equiv - h \gamma_\mu h^{-1} \, .
\end{equation}
Finally, the actual algebra is taken as
\begin{equation}
 \left\{ \gamma_\mu, \gamma_\nu \right\} = 2 \eta_{\mu\nu} \, .
\end{equation}
With these conventions, one can show that the effective action as written in \eqref{eq:GNansatz}
is indeed real if all field-dependent functions are real.

Let us also write down all possible products of Dirac matrices:
\begin{align}
 \gamma_\mu \gamma_\nu &= \eta_{\mu\nu} \mathbbm{1} + \Sigma_{\mu\nu} \, , \notag \\
 \gamma_\rho \Sigma_{\mu\nu} &= \eta_{\mu\rho} \gamma_\nu - \eta_{\nu\rho} \gamma_\mu + \epsilon_{\rho\mu\nu} \overline{\gamma} \, , \notag \\
 \gamma_\mu \overline{\gamma} &= \frac{1}{2} \epsilon_{\mu\nu\rho} \Sigma^{\nu\rho} \, , \notag \\
 \Sigma_{\mu\nu} \gamma_\rho &= \eta_{\nu\rho} \gamma_\mu - \eta_{\mu\rho} \gamma_\nu + \epsilon_{\mu\nu\rho} \overline{\gamma} \, , \notag \\
 \Sigma_{\mu\nu} \overline{\gamma} &= -\epsilon_{\mu\nu\rho} \gamma^\rho \, , \notag \\
 \Sigma_{\mu\nu} \Sigma_{\alpha\beta} &= \Sigma_{\alpha\mu} \eta_{\beta\nu} + \Sigma_{\beta\nu} \eta_{\alpha\mu}
 - \Sigma_{\alpha\nu} \eta_{\beta\mu} - \Sigma_{\beta\mu} \eta_{\alpha\nu}  \notag \\
 &\quad- (\eta_{\mu\alpha} \eta_{\nu\beta} - \eta_{\mu\beta} \eta_{\nu\alpha}) \mathbbm{1} \, , \notag \\
 \overline{\gamma} \, \Sigma_{\mu\nu} &= -\epsilon_{\mu\nu\rho} \gamma^\rho \, , \notag \\
 \overline{\gamma} \, \gamma_\mu &= \frac{1}{2} \epsilon_{\mu\nu\rho} \Sigma^{\nu\rho} \, , \notag \\
 \overline{\gamma} \, \overline{\gamma} &= -\mathbbm{1} \, .
\end{align}
Thus, with our conventions, no explicit factors of $\mathbf i$ appear.

\bibliography{specbib}

\end{document}